\documentclass[english,nofootinbib,aps,prd,twocolumn]{revtex4}
\usepackage[T1]{fontenc}
\usepackage[latin9]{inputenc}
\usepackage{amstext}
\usepackage{babel}
\usepackage{amsfonts}
\usepackage{mathrsfs}
\usepackage{amsmath}
\usepackage{amssymb}
\usepackage{bm}
\usepackage{extarrows}
\usepackage{slashed}
\usepackage{isodateo}
\usepackage{xcolor}
\usepackage{graphicx}
\usepackage{isodateo}
\usepackage[CJKbookmarks=true, bookmarksnumbered=true,bookmarksopen=true]{hyperref}
\hypersetup{colorlinks,%
              linkcolor=[rgb]{0,0.2,0.6}, %
              citecolor=[rgb]{0,0.2,0.6}}

\makeatletter
\@ifundefined{textcolor}{}
{%
 \definecolor{BLACK}{gray}{0}
 \definecolor{WHITE}{gray}{1}
 \definecolor{RED}{rgb}{1,0,0}
 \definecolor{GREEN}{rgb}{0,1,0}
 \definecolor{BLUE}{rgb}{0,0,1}
 \definecolor{CYAN}{cmyk}{1,0,0,0}
 \definecolor{MAGENTA}{cmyk}{0,1,0,0}
 \definecolor{YELLOW}{cmyk}{0,0,1,0}
}

\def\bge{\begin{equation}}
\def\ede{\end{equation}}
\def\bga{\begin{aligned}}
\def\eda{\end{aligned}}
\newcommand{\beq}{\begin{equation}}
\newcommand{\eeq}{\end{equation}}
\newcommand{\bq}{\begin{equation}}
\newcommand{\eq}{\end{equation}}
\newcommand{\ba}{\begin{array}}
\newcommand{\ea}{\end{array}}
\newcommand{\beqa}{\begin{eqnarray}}
\newcommand{\eeqa}{\end{eqnarray}}
\newcommand{\beqs}{\begin{subequations}}
\newcommand{\eeqs}{\end{subequations}}

\def\n{\nonumber}

\def\dis{\displaystyle}

\def\({\left(}
\def\){\right)}

\def\leqq{\leqslant}

\def\End{\end{document}}

\def\d{\text{d}}
\def\ii{{\tt i}}

\def\KK{\mathbb{K}}
\def\al{\alpha}
\def\be{\beta}
\def\ga{\gamma}
\def\la{\lambda}
\def\ep{\epsilon}
\def\mtd{\tilde{m}}

\makeatother

\begin{document}

\title{Connecting Leptonic Unitarity Triangle to Neutrino Oscillation \\
       with CP Violation in Vacuum and in Matter}

\author{{Hong-Jian He}\,$^{a,b}$ ~and~ {Xun-Jie Xu}\,$^{a,c}$}

\affiliation{$^a$Institute of Modern Physics and Center for High Energy Physics, Tsinghua
University, Beijing 100084, China
\\
$^b$Center for High Energy Physics, Peking University, Beijing 100871, China
\\
$^c$Max-Planck-Institut f\"{u}r Kernphysik, Postfach 103980, D-69029 Heidelberg, Germany
\\
(\,hjhe@tsinghua.edu.cn and xunjie.xu@gmail.com\,)}


\begin{abstract}
Leptonic unitarity triangle (LUT) provides fundamental means to geometrically describe
CP violation in neutrino oscillation. In this work, we use LUT to present a new geometrical
interpretation of the vacuum oscillation probability, and derive a compact
new oscillation formula in terms of only 3 independent parameters of the corresponding LUT.
Then, we systematically study matter effects in the geometrical
formulation of neutrino oscillation with CP violation. Including nontrivial matter
effects, we derive a very compact new oscillation formula by using the LUT
formulation. We further demonstrate that this geometrical formula
holds well for applications to neutrino oscillations in matter,
including the long baseline experiments T2K, MINOS, NO$\nu$A, and DUNE.
\\[2mm]
\hfill Phys.\ Rev.\ D\,(2016) in press [arXiv:1606.04054]
\end{abstract}
\maketitle

\section{\hspace*{-2.5mm}$\text{Introduction}$}
\label{sec:1}

Discovering leptonic CP violation poses a major challenge to particle physics today,
and may uncover the origin of matter-antimatter asymmetry in the Universe\,\cite{LEPG}.
Unitarity triangles provide the unique geometrical description of CP violations
via $3\!\times\!3$ unitary matrix.
They have played a vital role for studying CP violation of
Cabibbo-Kobayashi-Maskawa (CKM) mixings in the quark sector \cite{CKM}.
So far various neutrino oscillation experiments have been trying to precisely measure
Pontecorvo-Maki-Nakagawa-Sakata (PMNS) mixings for the lepton-neutrino sector \cite{PMNS}.
Leptonic unitarity triangles (LUT) provide a fundamental means
to probe the leptonic CP violation, complementary to the usual method
of measuring the CP asymmetry of neutrino oscillations,
$\,P[\nu_\ell^{}\!\to\!\nu_{\ell'}^{}] -
   P[\bar{\nu}_\ell^{}\!\to\!\bar{\nu}_{\ell'}^{}]\,$
($\ell\neq \ell'$) \cite{CP-AS,PDG2014}. Some LUT studies appeared
in the recent literature \cite{He:2013rba,LUTx,Xu:2014via,He:2015xha}.

\vspace*{1mm}
In Ref.\,\cite{He:2013rba}, we found that LUT is directly connected to
neutrino oscillations in vacuum.
We proved \cite{He:2013rba} that the LUT angles
exactly act as the CP-phase shifts of neutrino oscillations.
We proved \cite{He:2013rba} that vacuum oscillation only depends on
\,3\, independent geometrical parameters of the corresponding LUT.
Because matter effects \cite{MSWW,MSWMS}
in many current and future long baseline (LBL) oscillation experiments (such
as T2K\,\cite{Abe:2013hdq}, MINOS\,\cite{Adamson:2011ig}, NO$\nu$A\,\cite{Ayres:2004js},
and DUNE\,\cite{DUNE}) are non-negligible,
it is important to develop our geometrical LUT formulation
for including nontrivial matter effects.

\vspace*{1mm}

In this work, we construct a new unified geometrical LUT formulation for neutrino
oscillations in vacuum and in matter, and study its applications.
In Sec.\,\ref{sec:2},
we present a new geometrical LUT formulation
to dynamically describe how 3-neutrino system oscillates in vacuum.
From this, we derive a new compact oscillation formula,
manifestly in terms of only 3 independent
parameters of the corresponding LUT. In Sec.\,\ref{sec:3}, we systematically study the
LUT formulation for neutrino oscillations in matter. We derive an approximate analytical
LUT formula including matter effects, and further analyze its accuracy for the current
and future long baseline oscillation experiments, in Sec.\,\ref{sec:The-Accuracy-of} 
and Appendix\,\ref{app:A}-\ref{app:B}. 
Finally, we conclude in Sec.\,\ref{sec:conclusion}.

\vspace*{-2mm}
\section{\hspace*{-2.5mm}$\text{Geometrical Formulation of}$ \hspace*{4mm}
$\text{Neutrino Oscillation in Vacuum}$}
\label{sec:convention-and-basic}
\label{sec:2}
\vspace*{-2mm}

\begin{figure}
\centering
\includegraphics[width=8.1cm]{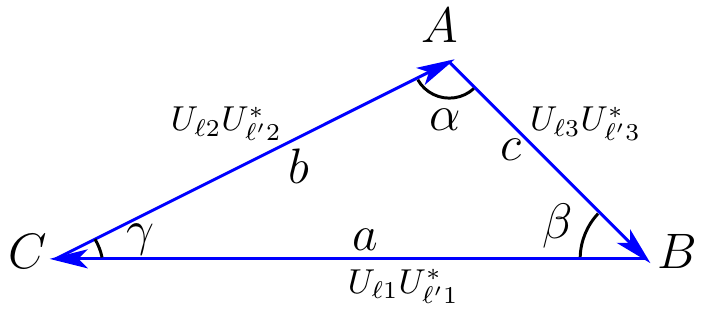}
\protect
\caption{The leptonic unitarity triangle (LUT), where $\,\ell\neq\ell'$,\,
\,$(a,\,b,\,c)$\, denote lengths of the three sides, and
\,$(\alpha,\,\beta,\,\gamma)$\, represent the three angles.
}
\label{fig:LUT}
\label{fig:1}
\end{figure}

From the unitarity of PMNS matrix, $\,U^{\dag}U=UU^{\dag}=1\,$,\,
we have two sets of conditions,
$\,\sum_j^{}U_{\ell j}^{}U_{\ell'j}^* = 0\,$ with $\,\ell\neq\ell'\,$
(forming the row triangles or ``Dirac triangles''),
and $\,\sum_\ell^{}U_{\ell j}^*U_{\ell j'}^{}=0\,$ with $\,j\neq j'\,$
(forming the column triangles or ``Majorana triangles'').
For the flavor neutrino oscillations, we consider the Dirac triangles
($\ell\neq\ell'$),
\beqa
U_{\ell1}^{}U_{\ell'1}^{*}+U_{\ell2}^{}U_{\ell'2}^{*}+U_{\ell3}^{}U_{\ell'3}^{*}
\,=\, 0\,, 
\label{eq:0530}
\eeqa
as shown in Fig.\ref{fig:1}.

\vspace*{1mm}

The sides and angles of each LUT \eqref{eq:0530} can be defined,
\beqs
\label{eq:abc-abg}
\beqa
(a,b,c) &\!=\!&
(|U_{\ell1}^{}U_{\ell'1}^{*}|,\, |U_{\ell2}^{}U_{\ell'2}^{*}|,\, |U_{\ell3}^{}U_{\ell'3}^{*}|),
\label{eq:1226-1}
\\[1.5mm]
(\alpha,\beta,\gamma) &\!=\!&
\arg\!\(\!-\frac{U_{\ell3}^{}U_{\ell'3}^{*}}{U_{\ell2}^{}U_{\ell'2}^{*}},
-\frac{U_{\ell1}^{}U_{\ell'1}^{*}}{U_{\ell3}^{}U_{\ell'3}^{*}},
-\frac{U_{\ell2}^{}U_{\ell'2}^{*}}{U_{\ell1}^{}U_{\ell'1}^{*}}\!\)\!.~~~~~~~
\label{eq:1226}
\eeqa
\eeqs
In Ref.\,\cite{He:2013rba}, we proved that the conventional neutrino oscillation probability
in vacuum\,\cite{CP-AS}\cite{PDG2014} can be fully expressed in terms of the sides
$(a,b,c)$ and angles $(\al,\be,\ga)$
of the corresponding LUT, among which only 3 are independent.

\vspace*{1mm}

In the following, we propose a new geometrical approach. With this,
we will derive a new formula of vacuum oscillations, which {\it manifestly contains
only 3 independent parameters of the LUT} for each given channel, say \,$(b,\,c,\,\alpha)$,\,
and takes a very compact form.

\vspace*{1mm}

In the standard formulation, the neutrino oscillation in space
may be described by the following Schr\"{o}dinger-like
evolution equation in flavor basis,
\beqa
{\ii}\frac{\d}{\d L}|\nu(L)\rangle =H|\nu(L)\rangle ,
\label{eq:1228}
\eeqa
where $\,H\,$ is the effective Hamiltonian and $|\nu(L)\rangle$ denotes
the flavor state of the flying neutrino at a distance $\,L\,$ from
the source. In vacuum, we can write the effective Hamiltonian $\,H_0^{}$\,
in the following matrix form,
\beqa
H_{0}^{} =
\frac{1}{\,2E\,}U\!\left(\!
\begin{array}{ccc}
m_{1}^{2}\\
 & m_{2}^{2}\\
 &  & m_{3}^{2}
\end{array}\!\right)\!U^{\dagger} .
\label{eq:1227-1-1}
\label{eq:H0}
\eeqa
Solving Eq.\eqref{eq:1228} gives,
$\,|\nu(L)\rangle =e^{-\ii H_{0}^{}L}|\nu(0)\rangle$.\,
So, the transition amplitude of
$\,\nu_{\ell}^{}\rightarrow\nu_{\ell'}^{}\,$  takes the form,
$\,A_{\ell\rightarrow\ell'}^{}=\sum_{j}U_{\ell j}^{}U_{\ell'j}^{*}e^{\ii 2\Delta_{j}^{}}$,\,
where $\,\Delta_{j}^{}\!\equiv m_j^2L/(4E)$.\,
Thus, we deduce the oscillation probability
$\,P_{\ell\rightarrow\ell'}^{}=|A_{\ell\rightarrow\ell'}^{}|^{2}$\,
as,
\begin{eqnarray}
\hspace*{-2mm}
P_{\ell\rightarrow\ell'}^{} & \!\!=\! &
|U_{\ell1}^{}U_{\ell'1}^{*}e^{\ii 2\Delta_{1}^{}}
\!+U_{\ell2}^{}U_{\ell'2}^{*}e^{\ii 2\Delta_{2}^{}}
\!+U_{\ell3}^{}U_{\ell'3}^{*}e^{\ii 2\Delta_{3}^{}}|^{2}
\nonumber
\\[1.5mm]
\hspace*{-2mm}
& \!\!=\! &
|a+be^{\ii (\gamma -\pi)}e^{\ii 2\Delta_{21}^{}}
\!+ce^{\ii (\pi -\beta)}e^{\ii 2\Delta_{31}^{}}|^{2} ,
\label{eq:1023}
\end{eqnarray}
where in the second row we have used Eq.\eqref{eq:abc-abg},
and $\,\Delta_{jk}^{}\equiv {\Delta m_{jk}^{2}L}/(4E)\,$ with
$\,\Delta m_{jk}^{2}=m_j^2-m_k^2$\, ($j,k=1,2,3$).

\begin{figure}[t]
\centering
\includegraphics[width=8.2cm]{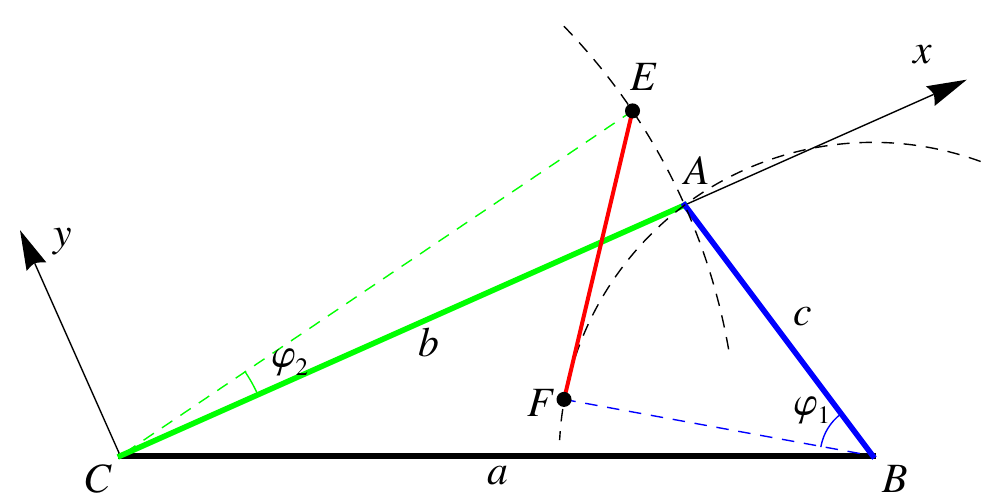}
\protect
\caption{New geometrical presentation of neutrino oscillation, where the angles
$\,\angle ABF =\varphi_{1}^{}$\, and \,$\angle ACE =\varphi_{2}^{}\,$ are evolving
phases $\,\varphi_{1}^{}=L\Delta m_{31}^{2}/(2E)$\, and
$\,\varphi_{2}^{}=L\Delta m_{21}^{2}/(2E)$\, in Eq.\,\eqref{eq:1109-7}.
The squared-distance $|EF|^2$ just gives the oscillation probability
\eqref{eq:1023} via Eqs.\eqref{eq:AB^2=P} and \eqref{eq:1021-2}.
If the triangle is closed, $|EF|=0$ and the probability vanishes. When $L$ increases,
the Point-E and Point-F will circle around the corresponding dashed arcs, and the distance
$|EF|$ oscillates. This means that the transition probability oscillates.}
\label{fig:geom}
\label{fig:2}
\end{figure}

\vspace*{1mm}

We inspect Eq.(\ref{eq:1023}) and find a new way to demonstrate its geometry graphically.
For $\,L/E=0$\,,\,  we have $\,P_{\ell\rightarrow\ell'}^{}=0\,$ and
$\,\Delta_{21}^{}=\Delta_{31}^{}=0$\,.\,
Hence, under $\,L/E=0$\,,\, Eq.\eqref{eq:1023} reduces to
\beqa
a+b\,e^{\ii (\gamma -\pi)}+c\,e^{\ii (\pi -\beta)} =\, 0 \,.\,
\label{eq:L/E=0}
\eeqa
This just corresponds to the geometry of the LUT $\bigtriangleup ABC$
shown in Fig.\,\ref{fig:1}.
We re-present this picture in Fig.\,\ref{fig:2}, where we have
\,$(|BC|,\,|CA|,\,|AB|)=(a,\,b,\,c)$,\, and
$(\angle CAB,\,\angle ABC,\,\angle BCA)=(\al,\,\be,\,\ga)$.\,
We see that the triangle geometry
$\,\overrightarrow{BC}+\overrightarrow{CA}+\!\overrightarrow{AB}=0\,$
just gives the equality \eqref{eq:L/E=0}.

\vspace*{1mm}

The generical case of $\,L/E\neq 0$\, has nonzero oscillation factors
$e^{\ii 2\Delta_{21}^{}}$ and $e^{\ii 2\Delta_{31}^{}}$.\,
This will modify the equality \eqref{eq:L/E=0},
in which $\,e^{\ii (\gamma-\pi)}$\, is replaced by
$\,e^{\ii (\gamma-\pi)}e^{\ii 2\Delta_{21}^{}}$\,,\,
and $\,e^{\ii (\pi-\beta)}$\, by \,$e^{\ii(\pi -\beta)}e^{\ii 2\Delta_{31}^{}}$,\,
causing the nonzero probability \eqref{eq:1023}.
Geometrically, the phase factors
$\,e^{\ii 2\Delta_{21}^{}}\,$ and
$\,e^{\ii 2\Delta_{31}^{}}$\, will change orientations of vectors $\overrightarrow{CA}$
and $\overrightarrow{BA}$ by holding their lengths.
This will rotate $\overrightarrow{CA}$ to $\overrightarrow{CE}$,\,
and $\overrightarrow{BA}$ to $\overrightarrow{BF}$, both counterclockwise.
Denoting the angles $\,\angle ABF= \varphi_{1}^{}\,$ and
$\,\angle ACE =\varphi_{2}^{}$,\, we have the following relations,
\beqa
\varphi_{1}^{}=2\Delta_{31}^{},~~~~~
\varphi_{2}^{}=2\Delta_{21}^{}.
\label{eq:1109-7}
\eeqa
This shows that the triangle is unfolded to become a quadrangle $ECBF$,
and the side $EF$ just equals the amplitude,
\beqa
\overrightarrow{EF} \,=\,
a+be^{\ii (\gamma-\pi)}e^{\ii 2\Delta_{21}^{}}
 +ce^{\ii (\pi-\beta)}e^{\ii 2\Delta_{31}^{}}.\,
\eeqa
Comparing this with Eq.\,\eqref{eq:1023}, we conclude that 
\beqa
\label{eq:AB^2=P}
|EF|^{2} \,=\, P_{\ell\rightarrow\ell'}^{}\,,
\eeqa
just gives the oscillation probability. When the quadrangle $ECBF$
reduces to a closed triangle $\bigtriangleup ABC$,
the oscillation probability would vanish.
When $L/E$ increases, the Point-$E$ and Point-$F$
in Fig.\,\ref{fig:geom} will circle around the corresponding dashed arcs.
Thus, the distance $|EF|$ oscillates, and its square $|EF|^2$ exactly equals
the oscillation probability \eqref{eq:1023} via Eq.\eqref{eq:AB^2=P}.
Hence, we have demonstrated that {\it Fig.\,\ref{fig:2} and Eq.\eqref{eq:AB^2=P}
give a new geometrical presentation of neutrino oscillations in vacuum.}

\vspace*{1mm}

We can directly compute the oscillation probability
by using the above geometrical formulation.
As will be shown below, it is striking
that using this geometrical formulation, we can derive
a very compact new formula of neutrino oscillations, {\it manifestly in
terms of only 3 LUT parameters.}
Without losing generality, we assign
$\,\overrightarrow{CA}\,$ as the x-axis and its orthogonal direction as y-axis.
Thus, we can derive the following coordinates for points $E$ and $F$
in the x-y plane,
\beqa
\hspace*{-4mm}
&E\!:&
(b\cos\varphi_{2}^{},\,b\sin\varphi_{2}^{}),
\label{eq:1023-1}
\nonumber
\\[-2mm]
\hspace*{-4mm}
\\[-1mm]
\hspace*{-4mm}
&F\!:&
\( b\!-\!2c\sin\!\frac{\varphi_1^{}}{2}\cos\!\angle FAC,\,
-2c\sin\!\frac{\varphi_1^{}}{2}\sin\!\angle FAC\)\!,~~~~~~
\label{eq:1023-2}
\nonumber
\eeqa
where the angle $\,\angle FAC \,=\, \alpha+\frac{1}{2}(\varphi_1^{}\!-\!\pi)$.\,
With these, we compute the length of the line segment $EF$ as
\begin{eqnarray}
|EF|^{2} & = &
 4c^{2}\!\sin^{2}\!\frac{\varphi_{1}^{}}{2\,}
 + 4b^{2}\!\sin^{2}\!\frac{\varphi_{2}^{}}{2\,}
\nonumber \\[0.6mm]
&& -8bc\sin\!\frac{\varphi_{1}^{}}{2}
   \sin\!\frac{\varphi_{2}^{}}{2}
   \cos\!\(\!\frac{\varphi_{1}^{}\!\!-\!\varphi_{2}^{}}{2}+\alpha\!\) \!.
\hspace*{12mm}
\label{eq:1023-6}
\end{eqnarray}
Using Eqs.\eqref{eq:1109-7}, \eqref{eq:AB^2=P}, and \eqref{eq:1023-6},
we derive an elegant and very compact new formula of vacuum oscillations,
\begin{eqnarray}
P_{\ell\rightarrow\ell'}^{} & \!\!=\! &
4c^{2}\sin^{2}\!\Delta
\nonumber\\[1.5mm]
&&
 -8bc\sin\!\Delta\sin\epsilon\Delta
 \cos[(1\!-\epsilon)\Delta +\alpha]
 \hspace*{12mm}
\label{eq:1021-2}
\\[1.5mm]
&& + 4b^{2}\sin^{2}\!\epsilon\Delta \,,
\nonumber
\end{eqnarray}
where we have defined,
\beqa
\Delta\equiv \Delta_{31}^{}=\frac{\,\Delta m_{31}^{2}L\,}{4E} ,
~~~~
\epsilon\equiv \frac{\Delta_{21}^{}}{\Delta_{31}^{}}
=\frac{\,\Delta m_{21}^{2}\,}{\Delta m_{31}^{2}} \,.~~~~
\label{eq:1110-1}
\eeqa
The anti-neutrino oscillation probability
$\,P_{\bar{\ell}\rightarrow\bar{\ell}'}^{}\,$
can be obtained from Eq.\eqref{eq:1021-2} under the replacement
$\,\,\alpha\to -\alpha\,$.\,
The new oscillation formula \eqref{eq:1021-2}
invokes {\it only 3 independent geometrical parameters
$\,(b,\,c,\,\alpha)$\, of the corresponding LUT,}
while the other 3 non-independent parameters
$\,(a,\,\beta,\,\gamma)$\,
have been explicitly removed in Eq.\eqref{eq:1021-2}.
Furthermore, this explicitly proves that the 4 PMNS-parameters $(\theta_{13}^{},\,\theta_{23}^{},\,\theta_{12}^{},\,\delta)$
could enter the oscillation probability \eqref{eq:1021-2}
{\it only via their 3 independent combinations
in terms of the geometrical parameters of LUT,}
such as $(b,\,c,\,\alpha)$.\,
Note that Eq.(\ref{eq:1021-2}) makes {\it no approximation.}
But it may be regarded as a Taylor expansion
in terms of $\,\sin\epsilon\Delta\,$ or $\,\ep\,$,\,
which is small due to
$\,\epsilon\approx 0.03\,$ \cite{Gfit}\cite{Gfit2}
and $\,\Delta\sim O(1)\,$
for all known accelerator oscillation experiments \cite{Abe:2013hdq}-\cite{DUNE}.
In Eq.(\ref{eq:1021-2}), the first row is of $O(\ep^0)$,\,
serving as the leading order (LO).
The second and third rows,
of $\,O(\epsilon^1)\,$ and $\,O(\epsilon^2)$,\, belong to the
next-to-leading order (NLO) and next-to-next-to-leading order (NNLO), respectively.
No other higher order terms exist because
Eq.(\ref{eq:1021-2}) is exact.

\vspace*{1mm}

Using our new Eq.\eqref{eq:1021-2}, we can rederive the oscillation
CP asymmetry
$\,A_{\text{CP}}^{\ell\ell'}
   =P_{\ell\rightarrow\ell'}^{} \!-\! P_{\bar{\ell}\rightarrow\bar{\ell}'}^{}\,$,
\beqa
\label{eq:A-CP}
A_{\text{CP}}^{\ell\ell'}
&\!\!=\!&
32\,S_{\triangle}^{}\!\sin\!\Delta\sin\ep\Delta\sin(1\!-\ep)\Delta
\\[1mm]
&\!\!=\!&
4J\!\(\sin\!2\Delta_{21}^{} \!\!+ \sin\!2\Delta_{13}^{} \!+\sin\!2\Delta_{32}^{}\)
\!,~~~~~~~~~
\nonumber
\eeqa
where $\,(\Delta,\,\ep )$\, are defined in Eq.\eqref{eq:1110-1},
and the Jarlskog invariant $J$ \cite{J} equals twice of the LUT area,
$\,J=2 S_{\triangle}^{}=bc\sin\alpha$\,.\,
The last line of Eq.\eqref{eq:A-CP} agrees to
the conventional CP asymmetry formula\,\cite{PDG2014}.

\vspace*{2mm}

As a final remark, we consider the conventional vacuum oscillation
formula\,\cite{CP-AS}\cite{PDG2014},
\begin{eqnarray}
\label{eq:U-lj}
P_{\ell\rightarrow \ell'}^{}
&\!=\!& \sum_{j=1}^{3}|U_{\ell'j}^{}U_{\ell j}^{}|^{2}
\\[-1mm]
&& +\,2\sum_{j<k}|U_{\ell'j}^{}U_{\ell j}^{}U_{\ell k}^{}U_{\ell'k}^{}|
   \cos (2\Delta_{jk}^{}\!\mp \phi_{\ell'\ell;jk}^{})\,,
\hspace*{8mm}
\n
\end{eqnarray}
where the signs ``$\mp$'' correspond to $\,\nu_{\ell}^{}\,(\bar{\nu}_{\ell}^{})\,$
oscillations.
Eq.\eqref{eq:U-lj} contains the CP phase angle\,\cite{CP-AS}\cite{PDG2014},
%
$\,\phi_{\ell'\ell;jk}^{} \equiv
\arg (U_{\ell'j}^{}U_{\ell j}^{*}U_{\ell k}^{}U_{\ell'k}^{*})$.\,
%
As we proved in Ref.\,\cite{He:2013rba}, each CP-phase shift $\,\phi_{\ell'\ell;jk}^{}\,$
exactly equals the corresponding angle of the LUT (modulo $\pi$), i.e.,
$\,(\phi_{\ell'\ell;23}^{},\phi_{\ell'\ell;31}^{},\phi_{\ell'\ell;12}^{})
 = (\alpha,\beta,\gamma)+\pi\,$,\,
where the convention of each LUT angle $(\alpha,\beta,\gamma)$ in
Eq.\eqref{eq:1226} differs from that of \cite{He:2013rba} by a minus sign.
With this, we derived the vacuum oscillation probability
$\,P[\nu_{\ell}^{}\rightarrow\nu_{\ell'}^{}]$,\,  fully in terms
of the geometrical parameters of the corresponding LUT \cite{He:2013rba},
\begin{equation}
\begin{array}{ll}
P_{\ell\rightarrow\ell'}^{} =\!
& 4ab\sin(\Delta_{12}^{}\!\mp\gamma)\sin\Delta_{12}^{}
\\[1mm]
& +\,4bc\sin(\Delta_{23}^{}\!\mp\alpha)\sin\Delta_{23}^{}
\\[1mm]
& +\,4ca\sin(\Delta_{31}^{}\!\mp\beta)\sin\Delta_{31}^{} \,,~~~~
\end{array}
\label{eq:1022-1}
\end{equation}
according to the current convention of Eq.\eqref{eq:abc-abg}.
Although Eq.(\ref{eq:1022-1})
contains all 6 parameters ($a,\,b,\,c$) and $(\alpha,\,\beta,\,\gamma$) of the LUT,
only 3 are independent. Hence, if we choose
3 of them, say $(b,\,c,\,\alpha)$,\,
the remaining parameters $(a,\,\beta,\,\gamma)$
can all be expressed in terms of $(b,\,c,\,\alpha)$,
%
\beqa
\hspace*{-6mm}
&& a \,= \sqrt{b^{2}\!+c^{2}\!-\!2bc\cos\!\alpha\,} \,,
\label{eq:1022-3}
\nonumber\\[-2mm]  
\\[-1.5mm]
\hspace*{-6mm}
&&
\gamma \,=\, \arccos\!\(\!\!\frac{\,a^2\!+\! b^2\!-\!c^2\,}{2ab}\!\!\)\!, ~~~
\beta \,=\, \pi -(\alpha \!+\! \gamma)\,.~~~~~
\hspace*{5mm}
\nonumber
\eeqa
%
We could try to eliminate the non-independent parameters $(a,\,\beta,\,\gamma)$\,
by substituting Eq.(\ref{eq:1022-3}) into Eq.\eqref{eq:1022-1}.
But the resultant form is very complicated and lengthy.
Only after we obtain the new formula \eqref{eq:1021-2} by the current
geometrical approach [Fig.\ref{fig:1} and Eq.\eqref{eq:AB^2=P}],
we could use Eq.\eqref{eq:1021-2} as the final answer (guideline),
and eventually reduce Eq.\eqref{eq:1022-1} to Eq.\eqref{eq:1021-2}
after tedious derivations.
Our new formula \eqref{eq:1021-2} is important, because extending it we can
further successfully construct the LUT formulation of neutrino oscillations
including nontrivial matter effects, as we will present
in Sec.\,\ref{sec:3}-\ref{sec:4}.

\vspace*{-2mm}
\section{\hspace*{-2.5mm}$\text{Neutrino Oscillations in Matter}$
\hspace*{3mm} $\text{and Effective Leptonic Unitarity Triangle}$}
\label{sec:LUT-Deformation-in}
\label{sec:3}
\vspace*{-2mm}

Including matter effects requires to add the following new term $\,H_{i}^{}\,$
into the effective Hamiltonian $\,H$ which appears in the evolution equation
(\ref{eq:1228}),
\beqa
H_{i}^{} = \sqrt{2}G_F^{}N_{e}\!
\left(\!\begin{array}{ccc}
1\\
 & 0\\
 &  & 0
\end{array}\right) \!,
\label{eq:1227-2}
\label{eq:Hi}
\eeqa
where the electron density $\,N_e= \,(Z/\!A)\,\rho\,N_A^{}\,$,
with $\,\rho\,$ the matter density, $Z$ ($A$) the atomic number
(atomic mass number), and $N_A^{}$ the Avogadro constant.
Eq.\eqref{eq:1227-2} is for neutrino oscillations in matter,
and for anti-neutrino oscillations the matter term \eqref{eq:1227-2}
flips sign \cite{PDG2014}.
Including this matter term \eqref{eq:1227-2}, we need to solve
Eq.\eqref{eq:1228} with $H$ given by
\begin{equation}
H \,=\, H_{0}^{}+H_{i}^{} \,.
\label{eq:1227}
\end{equation}
For the current LBL experiments, neutrino beams only pass through
the crust of the Earth. So $N_{e}$ is well approximated as a constant.
Thus, we derive the solution of Eq.\eqref{eq:1228},\,
$|\nu(L)\rangle = e^{-\ii HL}|\nu(0)\rangle\,.$\,
The Hamiltonian $H_{0}^{}$ can be diagonalized by the PMNS matrix $U$,\, but $H$
cannot, i.e., $\,U^{\dagger}H_{0}^{}U\,$
is diagonal, but $\,U^{\dagger}HU=U^{\dagger}H_{0}^{}U + U^{\dagger}H_{i}^{}U\,$
is not. Hence, we need to rediagonalize $H\,$ by an effective
mixing matrix $\,U_{m}^{}(\equiv U \!+\delta U)\,$,\,
which results in the effective neutrino masses $\tilde{m}_i^{}$.
Thus, we have
\begin{equation}
H\,=\frac{1}{\,2E\,}U_m^{}\!
\left(\!\!\begin{array}{ccc}
\tilde{m}_{1}^{2}\\
 & \tilde{m}_{2}^{2}\\
 &  & \tilde{m}_{3}^{2}
\end{array}\!\right)\! U_{m}^{\dagger}\,.
\label{eq:1104}
\end{equation}
From the effective mixing matrix $\,U_{m}^{}\,$,\,
we can construct the effective leptonic unitarity triangles (ELUT),
in the same way as we did for analyzing the vacuum LUT in Sec.\,\ref{sec:3}.
When neutrino energy $E$ is very low,
$\,H_{0}\gg H_{i}^{}\,$
and $U_{m}^{}$ is fairly close to $U$.\,
Hence, in the limit
$\,E\to 0$\,,\, the ELUT simply reduce to the corresponding LUT.
When $E$ increases, ELUT gradually deviate from LUT
since $U_{m}^{}$ deviates from $U$.\,
Thus, the forms of ELUT will vary under the change of neutrino energy $E$\,.

\vspace*{1mm}

The oscillation formula in matter is obtained by just replacing the original
LUT parameters, say, $(b,c,\alpha)$, by the new ELUT parameters
$(b_{m}^{},c_{m}^{}, \alpha_{m}^{})$.\,
We make the same replacements for effective neutrino masses
in Eq.(\ref{eq:1104}).

\vspace*{1mm}

This means that the geometrical presentation of neutrino oscillations in
Fig.\,\ref{fig:geom} still holds after including the matter effects.
The only difference is to replace the vacuum LUT by
the ELUT in matter and the neutrino masses $\,m_{1,2,3}^{}\,$
by $\,\tilde{m}_{1,2,3}^{}\,$.\,
When a neutrino propagates in matter and its distance $L$ increases,
the Point-E and Point-F in Fig.\,\ref{fig:geom}
will circle around the corresponding arcs in the ELUT frame.
Then, the distance $|EF|$ oscillates, and $|EF|^2$
gives the oscillation probability in matter.
Hence, including matter effects into Eq.\eqref{eq:1021-2},
we deduce the oscillation formula,
\begin{eqnarray}
P_{\ell\rightarrow\ell'}^{}
& \!\!\!=\!\! & 4c_{m}^{2}\sin^{2}\!\Delta_{m}^{}
\nonumber \\[1.5mm]
&& -8b_{m}^{}c_{m}^{}\sin\!\Delta_{m}\sin(\epsilon_{m}^{}\Delta_{m}^{})
 \cos[(1\!-\!\epsilon_{m}^{})\Delta_{m}^{}\!\!+\!\alpha_{m}^{}]~~
 \nonumber \\[1.5mm]
&& +4b_{m}^{2}\sin^{2}\,\!(\epsilon_{m}^{}\Delta_{m}^{})\,,
 \label{eq:1021-2-1}
\end{eqnarray}
where parameters with subscripts ``$\thinspace{}_{m}$''
denote the corresponding effective parameters in matter.
For instance, $b_{m}^{}$ is the $b$-side of the ELUT
from $\,U_{m}^{}$  in Eq.\eqref{eq:1104}.
$(\Delta_{m}^{},\,\ep_m^{})$ are obtained from
$(\Delta,\,\ep)$ [cf.\ Eq.\eqref{eq:1110-1}]
under the replacements $\,m_{1,2,3}^{}\to \tilde{m}_{1,2,3}^{}\,$
[cf.\ Eq.(\ref{eq:1104})].

\vspace*{1mm}

Note that Eq.(\ref{eq:1021-2-1}) is an exact formula, and so far we have not
made any approximation.
When neutrino energies lie between
the solar resonance and atmospheric resonance,
$0.1\,\text{GeV}\lesssim E\lesssim3\,\text{GeV}$ \cite{Freund},
one has the matter density
$\,\rho\simeq 2.6\,\text{g}/\text{cm}^3\,$ \cite{T2K-PRD} for the earth crust,
and the averaged ratio $Z/A\simeq 1/2$,\,
where $Z$ and $A$ are the atomic number and mass number, respectively.
With these, we deduce the approximate relations after a
nontrivial and lengthy derivation,
%
\beqs
\label{eq:bcan-m}
\beqa
&&\hspace*{-3mm}
c_{m}^{} \simeq \frac{c}{\,1\!-\!n_E^{}}\,, ~~~~
b_{m}^{} \simeq \frac{\epsilon \hspace*{0.3mm}b}{\,n_E^{}\,}\,,~~~~
\alpha_{m}^{}\simeq \alpha \pm \pi \,, \hspace*{13mm}
\label{eq:0111}
\\ 
&&\hspace*{-3mm}
\epsilon_{m}^{} \simeq \frac{-n_E^{}}{\,1\!-\!n_E^{}}\,, ~~~~
\Delta_{m}^{} \simeq (1\!-\! n_E^{})\Delta \,,
\label{eq:0111-1}
\eeqa
%
where 
$\,n_{E}^{}\,$ is defined as
 \beqa
 n_E^{} = 2\sqrt{2}\,G_{\!F}^{}N_{e}E/\Delta m_{31}^{2} \,.
 \label{eq:0111-2}
 \eeqa
 \eeqs
For clarity, we will present the nontrivial derivation of Eq.\eqref{eq:bcan-m}
in Appendix\,\ref{app:A}.
These are important relations connecting the ELUT parameters in matter
to the corresponding LUT parameters in vacuum.
They allow us to use the vacuum LUT parameters to directly compute
the oscillation probability in matter. This makes our LUT formulation
applicable to the current and future LBL oscillation experiments
\cite{Abe:2013hdq}-\cite{DUNE}.
In the following Sec.\,\ref{sec:4} as well as Appendix\,\ref{app:B}, 
we will perform numerical analysis to
explicitly test the accuracy of the above matter formulas
(\ref{eq:1021-2-1})-(\ref{eq:bcan-m}), and discuss their validity.

\vspace*{-2mm}
\section{\hspace*{-2.5mm}$\text{Applications: Testing the Precision of}$ \hspace*{3mm}
$\text{Geometrical Oscillation Formula}$}
\label{sec:The-Accuracy-of}
\label{sec:4}
\vspace*{-2mm}

So far, most of the LBL experiments measure neutrino appearance via
the oscillation channel $\,\nu_{\mu}^{}\rightarrow\nu_{e}^{}$\,.\,
Using our general geometrical equation (\ref{eq:1021-2-1})
together with the approximate relations (\ref{eq:0111})-(\ref{eq:0111-1}),
we derive the following analytical LUT formula for the
appearance oscillation probability,
\beqa
&& \hspace*{-13mm}
P_{\text{LUT}}^{}(\nu_{\mu}^{}\!\!\rightarrow\!\nu_e^{}) \,=
\nonumber\\[0.5mm]
&& \hspace*{-5mm}
\dis
\frac{4c^{2}}{\,(1\!-\!n_E^{})^{2}\,}\!\sin^{2}[(1\!\!-\!n_{E}^{})\Delta]
+\frac{\,4\epsilon^2b^2\,}{n_{E}^{2}}\!
 \sin^{2}(n_E^{}\Delta )~~
\nonumber\\[1mm]
&&
\hspace*{-5mm}
\dis
-\frac{\,8\epsilon bc\sin[(1\!\!-\!n_E^{})\Delta]
 \sin(n_{E}^{}\Delta)\cos(\Delta\!+\!\alpha )\,}
 {n_{E}^{}(1\!-\!n_{E}^{})\,}.
\label{eq:1023-4}
\label{eq:LUTm-mue}
\eeqa
The anti-neutrino oscillation probability
$\,P_{\text{LUT}}^{}(\bar{\nu}_{\mu}^{}\!\!\rightarrow\!\bar{\nu}_e^{})\,$
is obtained from Eq.\eqref{eq:1023-4} under the replacement
$(\alpha,\,n_E^{})\to (-\alpha,\,-n_E^{})$.\,
We stress that the new formula \eqref{eq:1023-4}
{\it is fully expressed in terms of only 3 independent
parameters $\,(b,\,c,\,\alpha)$\, of the LUT,} and
{\it is manifestly rephasing invariant.}
We also note that the form of Eq.\eqref{eq:1023-4} holds for both neutrino-mass-orderings.
For the normal mass-ordering ($m_1^{}< m_2^{}\ll m_3^{}$),
$\,\Delta$\, and \,$\ep$\, are positive,
while for the inverted mass-ordering ($m_2^{}> m_1^{}\gg m_3^{}$),\,
they are both negative.

\vspace*{1mm}

\begin{figure}[t]
\centering
\includegraphics[width=8.2cm,height=5.5cm]{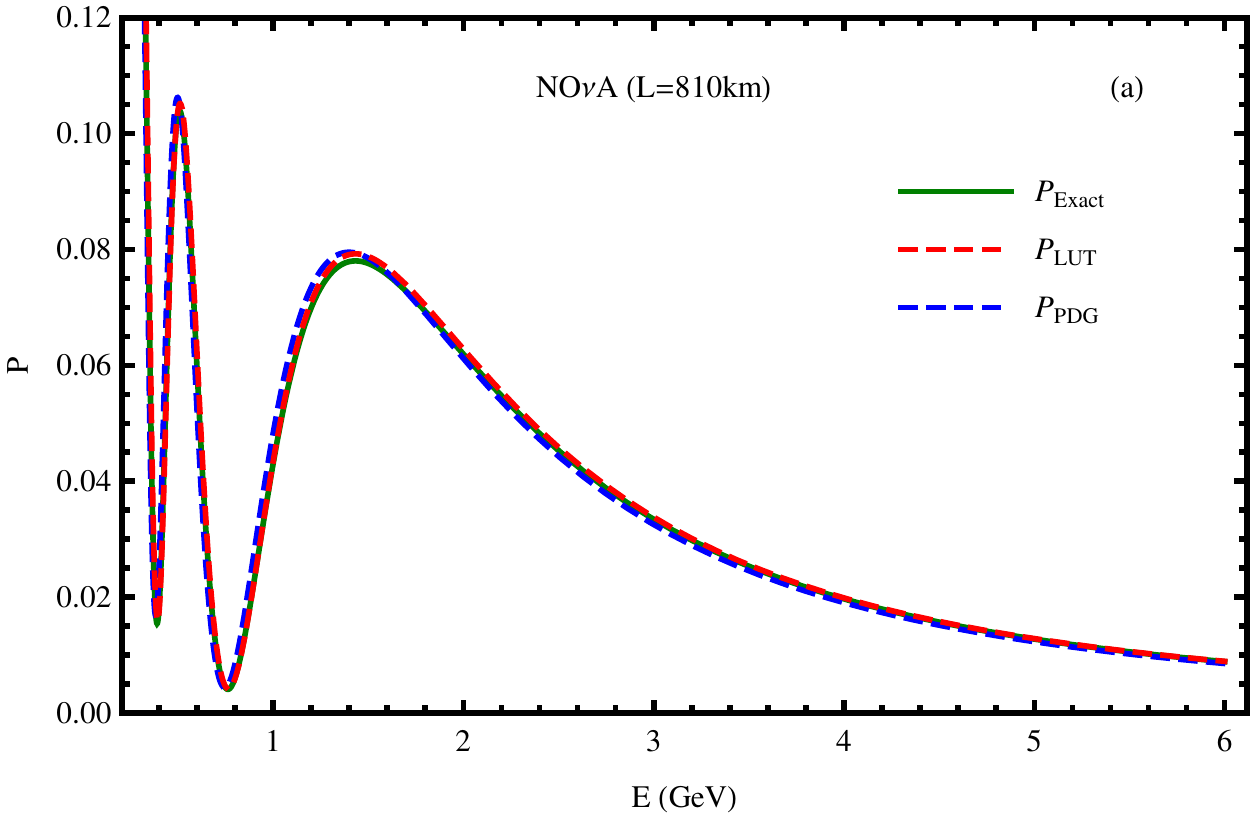}
\\
\hspace*{-3.4mm}
\includegraphics[width=8.5cm,height=5.5cm]{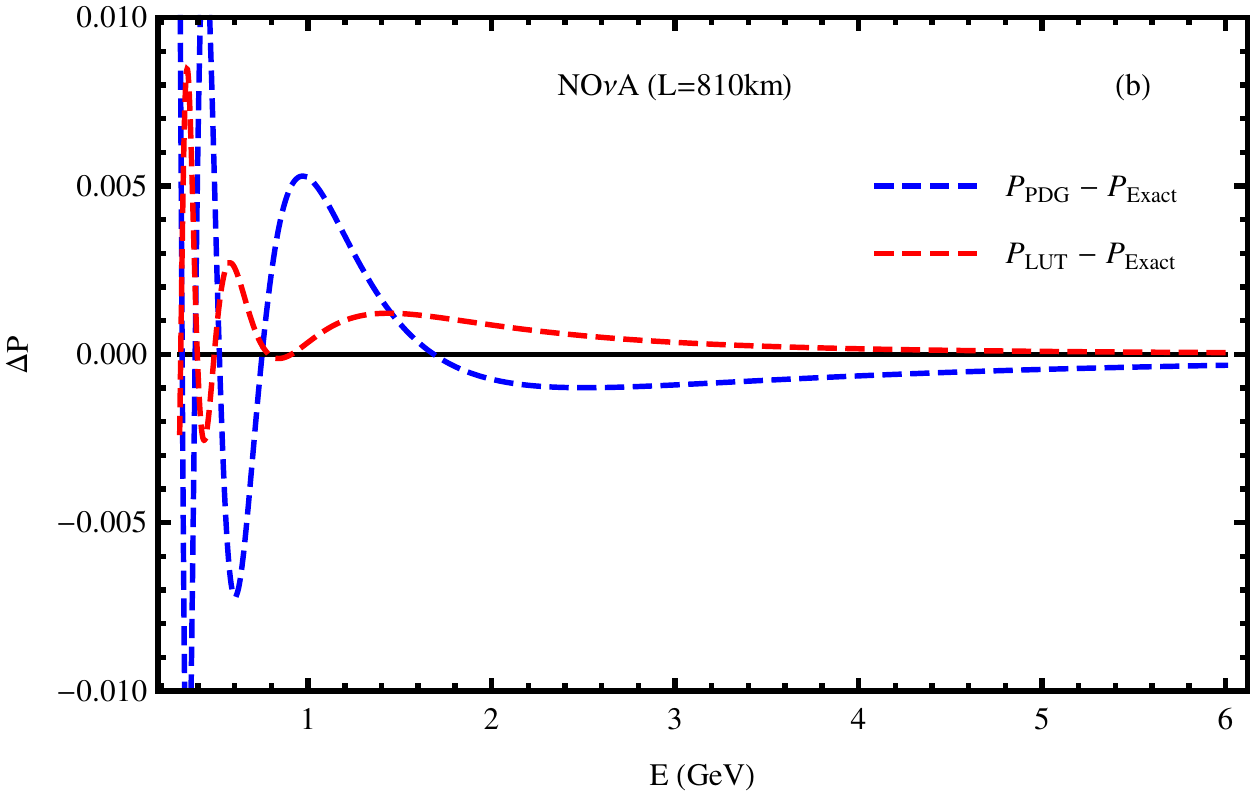}
\vspace*{-3mm}
\protect\caption{
Comparison of the approximate analytical oscillation formulae (\ref{eq:1023-4}) and
(\ref{eq:1023-37}) with the exact numerical result (green curve) for the case of NO$\nu$A
experiment ($L=810\,$km).\,
Eq.(\ref{eq:1023-4}) is plotted in red curve, and
Eq.(\ref{eq:1023-37}) is in blue curve.
Plot-(a) shows that both Eqs.(\ref{eq:1023-4})-(\ref{eq:1023-37})
are fairly accurate and their errors are negligible for practical use.
Plot-(b) depicts the differences
$\,\Delta P\!=\!P_{\text{LUT}}^{}-P\,$ (red curve) and
$\,\Delta P=P_{\text{PDG}}^{}\!-\!P\,$ (blue curve), showing that our
LUT formula (\ref{eq:1023-4}) is as accurate as
Eq.(\ref{eq:1023-37}).}
\label{fig:NOvA}
\label{fig:3}
\end{figure}

In the following, we analyze the accuracy of Eq.\eqref{eq:1023-4}
for practical applications.
We first compute the probability from Eq.(\ref{eq:1023-4}) and
compare it with the exact numerical result from solving the
neutrino evolution equation \eqref{eq:1228}.
We present this comparison in Fig.\,\ref{fig:3}(a)-(b)
for the on-going NO$\nu$A experiment with baseline $L=810$\,km.
In plot-(a), the red dashed curves depict the prediction $P_{\text{LUT}}^{}$ of
our LUT formula (\ref{eq:1023-4}),
and the green curve stands for the exact numerical result $P_{\text{Exact}}^{}$.\,
In Fig.\,\ref{fig:3}(b), we further present the difference
$\,\Delta P\!=\!P_{\text{LUT}}^{}-P_{\text{Exact}}^{}\,$ by red dashed curve.

\vspace*{1mm}

For the comparison in Fig.\,\ref{fig:3}, we further examine 
the approximate formula used by Particle Data Group
(PDG)\,\cite{PDG2014}\cite{Freund}\cite{Papp},
\begin{eqnarray}
 &  & \hspace*{-7mm}
 P_{\text{PDG}}^{}(\nu_{\mu}^{}\!\!\rightarrow\!\nu_{e}^{})=
\nonumber \\[1mm]
 &  & \frac{1}{\,(1\!-n_{E}^{})^{2}\,}\sin^{2}\!\theta_{a}^{}\sin^{2}\!2\theta_{x}^{}
      \sin^{2}[(1\!-\!n_{E}^{})\Delta]
      \nonumber \\[1mm]
 &  &
 -\frac{\epsilon}{\,n_E^{}(1\!-\!n_E^{})\,}\sin2\theta_{s}^{}\sin2\theta_{a}^{}
 \sin2\theta_{x}^{}\cos\theta_{x}^{}\sin\delta~~~~~
 \nonumber \\[1mm]
 &  & \hspace*{3mm}\times\sin\Delta\sin(n_E^{}\Delta)\sin[(1\!-\!n_E^{})\Delta]
 \nonumber \\[2mm]
 &  & +\frac{\epsilon}{\,n_E^{}(1\!-\!n_E^{})\,}
 \sin 2\theta_{s}^{}\sin 2\theta_{a}^{}\sin 2\theta_{x}^{}\cos\theta_{x}^{}\cos\delta
 \nonumber \\[1mm]
 &  & \hspace*{3mm}\times\cos\Delta\sin(n_E^{}\Delta)\sin[(1\!-\!n_E^{})\Delta]
 \nonumber \\[1mm]
 &  & +\frac{\epsilon^{2}}{n_{E}^{2}}
 \sin^2\! 2\theta_{s}^{}\cos^{2}\!\theta_{a}^{}
 \sin^{2}(n_{E}^{}\Delta)\,,
 \label{eq:1023-37}
 \label{eq:PDGm-mue}
\end{eqnarray}
where
$(\theta_s^{},\,\theta_a^{},\,\theta_x^{})
   \equiv (\theta_{12}^{},\,\theta_{23}^{},\,\theta_{13}^{})$
   and $\delta$ is CP angle.
Eq.\eqref{eq:1023-37} is widely adopted by LBL experiments for data analysis,
including the recent work of T2K\,\cite{T2K-PRD}.
(Some other approximate formulas using
the conventional PMNS parametrization appeared in the literature\,\cite{otherx}.)
Eq.\eqref{eq:1023-37} is much more complex than our LUT Eq.\,\eqref{eq:1023-4}.
For comparison, we plot the probability $\,P_{\text{PDG}}^{}\,$
by blue dashed curves in Figs.\,\ref{fig:3}(a).
We further depict the difference
$\,\Delta P=P_{\text{PDG}}^{}\!-\!P_{\text{Exact}}^{}\,$ (blue dashed curve)
in Figs.\,\ref{fig:3}(b). For illustrating the applications
of Eqs.(\ref{eq:1023-4})-(\ref{eq:1023-37}) in Fig.\,\ref{fig:3},
we have input central values of the current global fit \cite{Gfit} for
neutrino parameters under the normal mass-ordering.
We have also made similar comparisons under the inverted mass-ordering.

\vspace*{1mm}

Fig.\,\ref{fig:3} demonstrates that
for applications to LBL experiments (such as NO$\nu$A \cite{Ayres:2004js}),
our LUT formula \eqref{eq:1023-4} is very accurate and its error is negligible
for the current experimental precision. It shows that Eq.\eqref{eq:1023-4} is as precise as
or better than the widely-used PDG Eq.\eqref{eq:1023-37}.
Eq.\eqref{eq:1023-4} contains only 3 independent
LUT parameters $\,(b,\,c,\,\alpha)$,\, and is manifestly rephasing-invariant.
In contrast, Eq.\eqref{eq:1023-37} depends on all 4 PMNS-parameters
$(\theta_s,\,\theta_a,\,\theta_x,\,\delta)$.\,

\vspace*{1mm}

Note that Eq.\eqref{eq:1023-4} is derived from our independent new LUT approach
and stands on its own, even though Fig.\,\ref{fig:3}(a) shows that
Eqs.\eqref{eq:1023-4} and \eqref{eq:1023-37} are in main agreement.
We stress that
Eqs.\eqref{eq:1023-4} and \eqref{eq:1023-37} have their own advantages via two independent
formulations of $\nu$-oscillation; they are {\it complementary}
for studying different aspects of neutrino oscillations.
For current illustrations, we mainly consider the important on-going experiment
NO$\nu$A~($L=810$\,km)\,\cite{Ayres:2004js} as an example.
We have also reached similar conclusions
by analyzing other LBL experiments MINOS~($L=735$\,km)\,\cite{Adamson:2011ig}
and T2K~($L=295$\,km)\,\cite{T2K-PRD}, as well as
the planned future experiment DUNE~($L=1300$\,km)\,\cite{DUNE}.
For further justifications of our LUT matter formula \eqref{eq:LUTm-mue},
we will present explicit analyses for both
T2K and DUNE experiments in Appendix\,\ref{app:B},
covering a wide baseline range of $\,L=295-1300$\,km.

\vspace*{1mm}

In passing, we note that in principle,
both formulae (\ref{eq:1023-4}) and (\ref{eq:1023-37}) require
$\,\epsilon\ll n_{E}^{}\,$,\, which corresponds to a lower bound on neutrino energy,
\begin{equation}
E ~\gtrsim~ 0.34\textrm{GeV}\!\(\!\!
\frac{\Delta m_{21}^{2}}{\,7.6\!\times\!10^{-5}\textrm{eV}^{2}\,}
\frac{\,1.4\textrm{cm}^{-3}N_A^{}\,}{N_e}\!\!\) \!,~~~~
\label{eq:1225-3}
\end{equation}
as given in Ref.\,\cite{Freund} and updated by PDG\,\cite{PDG2014}
(cf.\ the note below Eq.(14.76) in Ref.\,\cite{PDG2014}).
For NO$\nu$A experiment, the selected neutrino energy
range is $\,1.5\text\,\text{GeV}\leqq E\leqq 2.7\,\text{GeV}$ \cite{Ayres:2004js},
which well obeys the lower bound \eqref{eq:1225-3}.
For the case of T2K experiment, it has neutrino energy range,
$\,0.1\text\,\text{GeV}\leqq E\leqq 1.2\,\text{GeV}$ \cite{Abe:2013hdq}.
So we may concern the validity
of our formula for $\,E=(0.1-0.34)\,$GeV.\,
Our derivation in Sec.\,\ref{sec:3} has made $\,\epsilon\,$ expansion, which
requires $\,\epsilon\ll n_E^{}\,$.\,
We note that the approximate Eq.\eqref{eq:0111} for $\,b_m^{}\,$
is singular in the limit $\,E\to 0$\, (which causes $\,n_E^{}\to 0\,$).
But our Eq.\eqref{eq:1023-4} is free from this singularity because
its $\,n_E^{}= 0\,$ poles are actually canceled
in the limit $\,n_E^{}\to 0\,$.\,
So, Eq.\eqref{eq:1023-4} still holds well around this limit.
Also,  a singularity $\,n_E^{}=1\,$ appears in
Eq.\eqref{eq:0111} for $\,c_m^{}\,$.\, Again, it is fully canceled in
Eq.\eqref{eq:1023-4}, and is harmless.
Note that the PDG Eq.\eqref{eq:1023-37} is also singularity-free in the limit
$\,n_E^{}\to 1$,\,  or, $\,n_E^{}\to 0\,$
[even though the perturbative expansion requires $\,\epsilon\ll n_E^{}\,$
and thus the bound \eqref{eq:1225-3}]. But exact numerical calculations have verified
that Eq.\eqref{eq:1023-37} remains fairly accurate below the bound (\ref{eq:1225-3}).
Hence, Eq.\eqref{eq:1023-37} was safely adopted
by T2K analysis \cite{Abe:2013hdq}.
Ref.\,\cite{Xu:2015kma} recently explained
why Eq.\eqref{eq:1023-37} still holds at energies below the bound (\ref{eq:1225-3}).
For our LUT Eq.\eqref{eq:1023-4}, we have demonstrated its validity for
various oscillation experiments by comparing it with the exact numerical results
in Fig.\,\ref{fig:3} and in Appendix\,\ref{app:B}.
We also expect similar reasons to explain the high numerical precision of
our LUT Eq.\eqref{eq:1023-4}, and will study the detail of this issue elsewhere.

\vspace*{-4mm}
\section{\hspace*{-2.5mm}$\text{Conclusions}$}
\label{sec:conclusion}
\label{sec:5}
\vspace*{-2mm}

Leptonic unitarity triangle (LUT) provides fundamental means
to geometrically describe CP violation in neutrino oscillations.
In this work, we presented a new unified geometrical formulation
for connecting the LUT to neutrino oscillations in vacuum and in matter.
We demonstrated that the dependence of the vacuum oscillation probability on
the PMNS mixing matrix can be fully reformulated in terms of
only 3 independent geometrical parameters of the corresponding LUT,
which are rephasing invariant.
We further constructed the geometrical formulation of oscillations in matter,
and derived a very compact and accurate new oscillation formula.

\vspace*{1mm}

In Sec.\,\ref{sec:2}, we proposed a new geometrical LUT formulation
of the dynamical 3-neutrino oscillations.
We proved that the vacuum oscillation probability can be derived by directly
computing the distance of two points circling around a vertex of the LUT,
as shown in Fig.\,\ref{fig:2} and given in Eqs.\eqref{eq:AB^2=P}(\ref{eq:1021-2}).
The formula \eqref{eq:1021-2} manifestly depends on only 3 independent parameters
of the corresponding LUT, and takes a much simpler form than
Eqs.(\ref{eq:1022-1})-\eqref{eq:1022-3} 
which we derived before\,\cite{He:2013rba}.
For neutrino oscillations in matter, we constructed the corresponding
Effective LUT (ELUT) in Sec.\,\ref{sec:3},
which is a deformed LUT by including matter effects.
Eqs.\eqref{eq:1021-2-1}-\eqref{eq:bcan-m}
presented a new geometrical oscillation formula including matter effects.
Note that Eqs.\eqref{eq:1021-2} and \eqref{eq:1021-2-1} exhibit LO+NLO+NNLO structure,
but hold exactly without approximation.
To analytically connect the ELUT parameters in Eq.\eqref{eq:1021-2-1}
to the vacuum LUT parameters, we deduced new relations
\eqref{eq:0111}-\eqref{eq:0111-1} under proper expansions, 
as shown in Appendix\,\ref{app:A}.
With these, we further derived a
very compact analytical formula (\ref{eq:1023-4}) in Sec.\,\ref{sec:4}.
We demonstrated that Eq.(\ref{eq:1023-4}) has high accuracy for
applications to long baseline experiments,
such as NO$\nu$A (Fig.\,\ref{fig:3}) and MINOS,
as well as T2K and DUNE (cf.\ Figs.\,\ref{fig:4}-\ref{fig:5}
in Appendix\,\ref{app:B}).
We showed that the numerical precision
of our LUT formula \eqref{eq:1023-4} is as good as (or better than)
the widely used PDG Eq.\eqref{eq:1023-37} \cite{PDG2014},
for the long baseline oscillation experiments T2K, MINOS, NO$\nu$A, and DUNE.

\appendix

\noindent
\section{\hspace*{0mm}$\textrm{Derivation of Matter Formula (22)}$}
\label{app:A}

In this Appendix, we present the highly nontrivial derivation of
the matter formula \eqref{eq:bcan-m},
shown at the end of Sec.\,III in the main text.

\vspace*{1mm}

Inspecting the effective Hamiltonian $\,H=H_0^{}+H_i^{}$\,
in Eqs.\eqref{eq:H0} and \eqref{eq:Hi},
we can separate out a diagonal term $\,(m_{1}^{2}/2E)\mathbb{I}\,$
and express $H$ as follows,
\beqa
\label{eq:H-IK}
H \,=\,
 \frac{m_{1}^{2}}{\,2E\,}\mathbb{I}
+\frac{\,\Delta m_{31}^{2}\,}{\,2E\,}U \KK\, U^{\dagger} ,
\eeqa
where $\,\mathbb{I}\,$ is the $3\times3$ unit matrix.
The dimensionless matrix $\,\KK\,$ takes the following convenient form,
and needs to be diagonalized,
\begin{equation}
\KK \,=\left(\!\begin{array}{ccc}
0\\
 & \epsilon\\
 &  & 1
\end{array}\!\right)+n_{E}^{}u_{e}^{}u_{e}^{\dag} \,,
\label{eq:1023-13}
\end{equation}
where $\,u_{e}^{\dag}$ is the first row of $\,U$.\, To be concrete, we
parametrize $U$ as
\beqa
U\!=\!\!\left(\!\!\!\!\begin{array}{ccc}
c_{s}^{}c_{x}^{} & s_{s}^{}c_{x}^{} & s_{x}^{}
\\[1.5mm]
-e^{-\ii\delta}s_{s}^{}c_{a}^{}\!\!-\!c_{s}^{}s_{a}^{}s_{x}^{}
& e^{-\ii\delta}c_{s}^{}c_{a}^{}\!\!-\!s_{s}^{}s_{a}^{}s_{x}^{} & s_{a}^{}c_{x}^{}
\\[1.5mm]
e^{-\ii\delta}s_{s}^{}s_{a}^{}\!\!-\!c_{s}^{}c_{a}^{}s_{x}^{}
& -e^{-\ii\delta}c_{s}^{}s_{a}^{}\!\!-\!s_{s}^{}c_{a}^{}s_{x}^{} & c_{a}^{}c_{x}^{}
\end{array}\!\!\right)\!\!,~~~~~
\label{eq:1105}
\eeqa
where we have used notations,
$\,(s_j^{},\,c_j^{})= (\sin\theta_j^{},\,\cos\theta_j^{})$,\,
and
$\,(\theta_s^{},\,\theta_a^{},\,\theta_x^{})
   = (\theta_{12}^{},\,\theta_{23}^{},\,\theta_{13}^{})\,$.\,
Thus, we have
\beqa
u_{e}^{} \,=\, (c_{s}^{}c_{x}^{},\,s_{s}^{}c_{x}^{},\,s_{x}^{})^{T} ,
\label{eq:1023-14}
\eeqa
where $\,u_{e}^{}\,$ is real under this convention, and thus
$\,u_{e}^{\dag}=u_{e}^{T}$.\,
Hence, we are actually going to diagonalize a real matrix $\,\KK\,$.\,
The final result of computing the ELUT does not depend on
the parametrization of $\,U$.\,
Note that Eq.(\ref{eq:1105}) can be obtained from the standard parametrization
of PMNS matrix ${U_{\!S}^{}}$ \cite{PDG2014} via simple rephasing,
\beqa
U=\,\text{diag}(1,\,e^{-\ii\delta}\!,\,e^{-\ii\delta})\,U_{\!S}^{}\,
  \text{diag}(1,\,1,\,e^{\ii\delta}) \,,~~~~
\label{eq:1105-1}
\eeqa
where $\,U_{\!S}^{}\,$ is given by \cite{PDG2014},
\beqa
U_{\!S}^{} \!=\!\left(\!\!\!\begin{array}{ccc}
c_{s}^{}c_{x}^{} & s_{s}^{}c_{x}^{} & e^{-\ii\delta}s_{x}^{}
\\[1.5mm]
-s_{s}^{}c_{a}^{}\!\!-\!e^{\ii\delta}c_{s}^{}s_{a}^{}s_{x}^{}
& c_{s}^{}c_{a}^{}\!\!-\!e^{\ii\delta}s_{s}^{}s_{a}^{}s_{x}^{}
& s_{a}^{}c_{x}^{}
\\[1.5mm]
s_{s}^{}s_{a}^{}\!\!-e^{\ii\delta}c_{s}^{}c_{a}^{}s_{x}^{}
& -c_{s}^{}s_{a}^{}\!\!-e^{\ii\delta}s_{s}^{}c_{a}^{}s_{x}^{} & c_{a}^{}c_{x}^{}
\end{array}\!\!\right)\!\!.~~~~~
\label{eq:1105-2}
\eeqa

\vspace*{1mm}

Using the following 1-2 rotation
\begin{equation}
O_{12}^{}\,=\left(\!\begin{array}{ccc}
c_{s} & -s_{s} & 0 \\[1mm]
s_{s} & c_{s} & 0  \\[1mm]
0 & 0 & 1
\end{array}\right) \!,~~~~~~
\label{eq:1023-15}
\end{equation}
we can rotate $u_{e}^{}$ into a vector containing only two non-zero elements,
\beqa
u_{e}^{\dag}O_{12}^{} \,=\, u_{e}^{T}O_{12}^{} \,=\,
(c_{x}^{},\, 0,\, s_{x}^{}) \,.
\label{eq:1023-16}
\eeqa
Then, we find that $\,n_{E}^{}u_{e}^{}u_{e}^{\dag}\,$
in Eq.(\ref{eq:1023-13})
will be rotated into a matrix having only two off-diagonal elements,
\beqa
n_{E}^{}O_{12}^{T}(u_{e}^{}u_{e}^{\dag})O_{12}^{}\,=\,
n_{E}^{}\!\left(\!\!\begin{array}{ccc}
c_{x}^{2} & 0 & s_{x}^{}c_{x}^{}
\\[1.5mm]
0 & 0 & 0
\\[1.2mm]
s_{x}^{}c_{x}^{} & 0 & s_{x}^{2}
\end{array}\!\!\right) \!.~~~~~
\label{eq:1105-3}
\eeqa

\vspace*{1mm}

For $\,\epsilon=0$\,,\, after the 1-2 rotation $O_{12}^{}$,
Eq.(\ref{eq:1023-13}) will be rotated to
\beqa
O_{12}^{T}\KK_{0}^{}O_{12}^{}\,=\left(\!\begin{array}{ccc}
 0\\
 & 0\\
 &  & 1
\end{array}\!\right)\!
+n_{E}^{}\!\left(\!\!\begin{array}{ccc}
c_{x}^{2} & 0 & s_{x}^{}c_{x}^{}
\\[1.5mm]
0 & 0 & 0\\[1.2mm]
s_{x}^{}c_{x}^{} & 0 & s_{x}^{2}
\end{array}\!\!\right) \!,~~~~~
\label{eq:1023-17}
\eeqa
where $\,\KK_{0}^{}\,$ is defined as, $\,\KK_{0}^{}=\KK |_{\epsilon=0}^{}$\,.\,
Thus, to diagonalize the matrix in Eq.\,(\ref{eq:1023-17}),
we just need an 1-3 rotation,
\begin{equation}
O_{13}^{}=\left(\!\!\begin{array}{ccc}
c_{\theta}^{} & 0 & ~s_{\theta}^{}
\\[1mm]
0 & 1 & ~0
\\[0.3mm]
-s_{\theta}^{} & 0 & ~c_{\theta}^{}
\end{array}\!\right) \!,
\label{eq:1023-18}
\end{equation}
where $(s_{\theta}^{},\,c_{\theta}^{}) = (\sin\theta,\,\cos\theta)$.\,
Under the rotation \eqref{eq:1023-18}, Eq.(\ref{eq:1023-16})
is transformed to
\beqa
u_{e}^{\dag}O_{12}^{}O_{13}^{} \,=\,
\( c_y^{},\,0,\, s_y^{}\)\,,
\label{eq:1023-19}
\eeqa
where $(s_{y}^{},\,c_{y}^{}) = (\sin\theta_y^{},\,\cos\theta_y^{})$,\,
and $\,\theta_y^{}\! = \theta + \theta_x^{}$\,.\,
Then, the matrix (\ref{eq:1023-17}) is rotated to
\begin{eqnarray}
 &  & \hspace*{-10mm}
O_{13}^{T}O_{12}^{T}\,\KK_0^{}\,O_{12}^{}O_{13}^{}
 \nonumber \\[2mm]
 & = & \left(\!\!\!\begin{array}{ccc}
s_{\theta}^{2} & 0 & -s_{\theta}^{}c_{\theta}^{}
\\[1.3mm]
0 & 0 & 0
\\[1mm]
-s_{\theta}^{}c_{\theta}^{} & 0 & c_{\theta}^{2}
\end{array}\!\!\right)
+ n_{E}^{}\!\left(\!\!\!\begin{array}{ccc}
c_{y}^{2} & 0 & s_{y}^{}c_{y}^{}
\\[1.5mm]
0 & 0 & 0\\[0.7mm]
s_{y}^{}c_{y}^{} & 0 & s_{y}^{2}
\end{array}\!\!\right)~~~~~
\nonumber
\\[2mm]
 & = & \left(\!\!\begin{array}{ccc}
\,\lambda_{-}^{} & 0 & 0\\
0 & 0 & 0\\
0 & 0 & \,\lambda_{+}^{}
\end{array}\!\right) \!,
\label{eq:1023-21}
\end{eqnarray}
where we have defined,
\beqa
\lambda_{-}^{} =\, s_{\theta}^{2}+n_{E}^{}c_{y}^{2}\,,~~~~~
\lambda_{+}^{} =\, c_{\theta}^{2}+n_{E}^{}s_{y}^{2}\,.
\label{eq:1105-4}
\eeqa
In the second equality of Eq.\eqref{eq:1023-21}, we have imposed
the following condition on the (1,3) and (3,1) elements to ensure
full diagonalization,
\beqa
n_{E}^{}s_{y}^{}c_{y}^{}-s_{\theta}^{}c_{\theta}^{} = 0 \,.~~~
\label{eq:1023-22}
\eeqa
This leads to
\beqa
\sin(2\theta) \,=\, n_{E}^{}\sin(2\theta_y^{}) \,.
\label{eq:1023-23}
\eeqa
Given the relation $\,\theta_{y}^{}\!=\theta+\theta_{x}^{}$,\,
we can solve $\,\theta$\, as a function of $\,\theta_x^{}\,$,
\beqa
\tan 2\theta \,=\,
\frac{\,n_E^{}\sin (2\theta_x^{})}
     {\,1\!-n_{E}^{}\cos (2\theta_x^{})\,} \,.
\label{eq:1023-24}
\eeqa
Hence, we have determined the 1-3 rotation
and diagonalized the matrix $\,\KK\,$ under
$\,\epsilon=0\,$ limit. The diagonalization matrix is $\,O_{12}^{}O_{13}^{}\,$,\,
and the effective mixing matrix in this case corresponds to
$\,U_{m0}^{} \equiv U_{m}^{}|_{\epsilon=0}^{}$\,,\, as given by
\begin{eqnarray}
U_{m0}^{} = U O_{12}^{}O_{13}^{}
 =  \!\left(\!\!\!\begin{array}{ccc}
c_{y}^{} & 0 & s_{y}^{}
\\[1mm]
-s_{y}^{}s_{a}^{} & c_{a}^{}e^{-\ii\delta} & c_{y}^{}s_{a}^{}
\\[1mm]
-s_{y}^{}c_{a}^{} & -s_{a}^{}e^{-\ii\delta} & c_{y}^{}c_{a}^{}
\end{array}\!\!\right)\!.~~~~~~
\label{eq:1023-25}
\end{eqnarray}
Hence, when $\,\epsilon=0$\,,\, the effective unitarity triangle extracted
from Eq.\,(\ref{eq:1023-25}) is actually a line for $\nu_e^{}\!-\nu_\mu^{}$\,
or $\,\nu_e^{}-\nu_\tau^{}$\, oscillations, since in either case the length of the
$b$ side vanishes,
\beqa
b_{m0}^{} = 0 \,.
\label{eq:1023-28}
\eeqa
The other two sides of this LUT have the same length,
\begin{eqnarray}
a_{m0}^{} \,=\, c_{m0}^{}
\,=\,
\frac{c}{\sqrt{(1\!-\!n_{E}^{})^{2}\!+\! 4n_E^{}s_{x}^{2}\,}\,}\,,
\label{eq:1109-5}
\end{eqnarray}
where $\,c = s_x^{}c_x^{}s_a^{}\,$ (or, $\,c = s_x^{}c_x^{}c_a^{}$)
is the length of $\,c\,$ side of the vacuum LUT
for $\nu_e^{}\!\!-\!\nu_\mu^{}$ (or, $\nu_e^{}\!\!-\!\nu_\tau^{}$) oscillations.
We have made use of Eqs.\eqref{eq:1023-23}-\eqref{eq:1023-24} for deriving
the formula \eqref{eq:1109-5}.

\vspace*{1mm}

Next, we compute the corrections from nonzero $\,\epsilon\,$.\,
For $\,\epsilon\neq0\,$,\, we can split the matrix $\,\KK\,$
in Eq.\eqref{eq:1023-13} as follows,
\beqa
\KK \,\equiv\, \KK_{0}^{} + \KK_{\epsilon} \,,
\label{eq:1023-29}
\eeqa
where
\beqa
\KK_0^{} = \!\left(\!\begin{array}{ccc}
0\\
& 0\\
&  & 1
\end{array}\!\right)\!
+ n_E^{}u_e^{}u_e^T ,
~~~~~
\KK_{\epsilon} = \!\left(\!\begin{array}{ccc}
0\\
& \epsilon\\
&  & 0
\end{array}\!\right)\!.~~~~~
\label{eq:1109-1}
\eeqa
Then, under the rotation $\,O_{12}^{}O_{13}^{}\,$,\,
the matrix $\,\KK_\epsilon\,$ transforms as
\begin{equation}
\hspace*{-1.5mm}
O_{13}^{T}O_{12}^{T}\,\KK_{\epsilon}O_{12}^{}O_{13}^{}
=\, \epsilon\!\left(\!\!\!\begin{array}{ccc}
s_{s}^{2} & s_{s}^{}c_{s}^{} & 0
\\[1mm]
s_{s}^{}c_{s}^{} & c_{s}^{2} & 0
\\[1mm]
0 & 0 & 0
\end{array}\!\right)\!
+\mathcal{O}(10^{-3}),~~~~
\label{eq:1023-30}
\end{equation}
where the corrections due to $O_{13}^{}$ rotation are suppressed by
$\,\epsilon\,\theta_{13}^{} = \mathcal{O}(10^{-3})$,\,
and are negligible in the final result.
Hence, under the rotations $\,O_{12}^{}O_{13}^{}\,$,\,
we have the matrix $\KK$ transform as
\beqa
\KK \rightarrow \!\left(\!\!\begin{array}{ccc}
\lambda_{-} & 0 & 0\\[0.6mm]
0 & 0 & 0\\[0.6mm]
0 & 0 & \lambda_{+}
\end{array}\!\!\right)\!\! +\epsilon\!\left(\!\!\!\begin{array}{ccc}
s_{s}^{2} & s_{s}c_{s} & 0
\\[1mm]
s_{s}c_{s} & c_{s}^{2} & 0
\\[1mm]
0 & 0 & 0
\end{array}\!\right)\!\! + \mathcal{O}(10^{-3}).~~~~~
\label{eq:1109-2}
\eeqa

\vspace*{1mm}

We can further diagonalize the right-hand-side of Eq.\eqref{eq:1109-2}
by a rotation $O'_{12}$\,,
\beqa
O'_{12}=\left(\begin{array}{ccc}
c_{r}^{} & -s_{r}^{} & 0 \\[1mm]
s_{r}^{} & c_{r}^{} & 0 \\[1mm]
0 & 0 & 1
\end{array}\right)\!,
\label{eq:1023-31}
\eeqa
where $(s_{r}^{},\,c_{r}^{})= (\sin\theta_r^{},\,\cos\theta_r^{})$.\,
Thus, we can determine the angle $\,\theta_r^{}\,$ as follows,
\beqa
\tan2\theta_{r}^{} =
\frac{\epsilon\sin(2\theta_s^{})}
     {\,\lambda_{-}^{}\!-\epsilon\cos(2\theta_s^{})\,} \,.
\label{eq:1023-24-1}
\eeqa
With these, we combine the above rotation $\,O'_{12}\,$
with $\,U_{m0}^{}\,$ in Eq.(\ref{eq:1023-25}),
and deduce the following rotation for full diagonalization,
\beqa
U_{m}^{} &\!\!=\!& U_{m0}^{}O'_{12}
\label{eq:1023-32}
\\[2mm]
&\!\!=\!&\!
\left(\!\!\!\begin{array}{ccc}
c_{y}^{}c_{r}^{} & -s_{r}^{}c_{y}^{} & s_{y}^{}
\\[2mm]
s_{r}^{}c_{a}^{}e^{-\ii\delta}\!\! -\! c_{r}^{}s_{a}^{}s_{y}^{}
& c_{r}^{}c_{a}^{}e^{-\ii\delta}\!\!+\! s_{r}^{}s_{y}^{}s_{a}^{}
& c_{y}^{}s_{a}^{}
\\[1.5mm]
\times & \times & \times
\end{array}\!\!\!\right) \!\!,~~~~~~
\nonumber
\eeqa
where $\times$ stands for the elements irrelevant to our current concern.
For the $\,\nu_e^{}\!-\nu_\mu^{}$\, LUT, using
Eqs.(\ref{eq:1023-32}) and (\ref{eq:1023-24-1}), we derive
\begin{eqnarray}
b_{m}^{} &=& s_{r}^{}c_{y}^{}
\left|c_{r}^{}c_{a}^{}e^{-\ii\delta} \!\!+s_{r}^{}s_{y}^{}s_{a}^{}\right|
\nonumber\\[1.5mm]
 & = &
 \frac{\epsilon s_{2s}^{}c_{a}^{}c_{y}^{}}
      {\,2\sqrt{(\lambda_{-}\!\!-\epsilon c_{2s}^{})^{2}\!+\!\epsilon^{2}s_{2s}^{2}\,}\,}
      [1\!+\mathcal{O}(\epsilon\theta_x^{})] \,,~~~~
\end{eqnarray}
where
$\,(s_{2s}^{},\,c_{2s}^{})= (\sin 2\theta_s^{},\,\cos 2\theta_s^{})$\,.\,
This gives a small but non-zero length for the $b$-side of the deformed
effective unitarity triangle,
\beqa
b_{m}^{} \simeq \frac{\epsilon}{\,\,\lambda_{-}^{}\,}b \,.
\label{eq:1023-33}
\eeqa
The length of $c$-side is not changed from the leading order result
because the last column of $\,U_m^{}\,$ equals that of $\,U_{m0}^{}$,\,
\beqa
c_{m}^{} \simeq c_{m0}^{} \,,
\label{eq:1109-4}
\eeqa
with $\,c_{m0}^{}\,$ given by Eq.\eqref{eq:1109-5}.
Since the current global fits of neutrino data\,\cite{Gfit}\cite{Gfit2}
restrict the $3\sigma$ range of
$\,\theta_{13}^{}\equiv\theta_x^{}\simeq 0.137-0.158 = \mathcal{O}(0.1)\ll 1\,$,\,
we see that $\,s_{y}^2\sim s_\theta^2\sim s_{x}^2 = \mathcal{O}(10^{-2})\,$
are fairly small. This applies to Eqs.(\ref{eq:1109-5}) and (\ref{eq:1105-4}).
Hence, we deduce the approximate relations,
$\,b_m^{}\simeq \epsilon\, b/n_E^{}\,$ and
$\,c_m^{}\simeq c\,/(1\!-n_E^{})\,$,\, which lead to Eq.\,\eqref{eq:0111}.

\vspace*{1mm}

From the definition of $\,\alpha\,$ in Eq.(2b),
and using the formulas \eqref{eq:1105} and \eqref{eq:1023-32},
we have $\,\alpha=-\arg(-U_{\mu2}^{})\,$
and $\,\alpha_{m}^{}=-\arg((U_{m})_{\mu2}^{})$,\,
where the expression of $\,\alpha_{m}^{}\,$ does not have a ``$-$'' sign
in front of $\,(U_{m})_{\mu2}^{}\,$ because it is canceled by the
negative sign in $\,(U_{m})_{e2}^{}$\,.\,
Ignoring $\,s_x^{},s_y^{}=\,\mathcal{O}(0.1)$,\,  we have
$\,\arg(U_{\mu2}^{})\simeq \arg((U_{m})_{\mu2}^{})$\,,\,
and thus $\,\alpha_{m}^{} \simeq \alpha \pm \pi\,$,\,
which reproduces the third relation of Eq.\eqref{eq:0111}.
Hence, the final ELUT is approximately given by Eq.\eqref{eq:0111}.
To derive Eq.\eqref{eq:0111-1},  we note that at $\,\mathcal{O}(\ep^0)\,$,\,
the eigenvalues of the matrix $\,\KK\,$ in Eq.\eqref{eq:1109-2} are
$(\la_-^{},\,0,\,\la_+^{})$.\, Accordingly, the effective Hamiltonian \eqref{eq:H-IK}
has three eigenvalues,
\beqa
\(\!
\frac{\,m_1^2\!+\!\Delta m_{31}^2\la_-^{}\,}{2E},\,
\frac{\,m_1^2\,}{2E},\,
\frac{\,m_1^2\!+\!\Delta m_{31}^2\la_+^{}\,}{2E}
\!\)\!,~~~~
\eeqa
which should equal the corresponding eigenvalues
\,$(\mtd_1^2,\,\mtd_2^2,\,\mtd_3^2)/(2E)$\, as
defined in Eq.\eqref{eq:1104}.
Hence, we deduce the effective mass-squared differences
at $\,\mathcal{O}(\ep^0)\,$,
\beqa
\Delta\tilde{m}_{31}^2 &\simeq&
(\lambda_{+}^{}\!\!-\!\lambda_{-}^{})\Delta m_{31}^2,\,~~~~
\nonumber\\[-1.5mm]
&& \\[-1.5mm]
\Delta\tilde{m}_{21}^2 &\simeq& 
(-\lambda_{-}^{})\Delta m_{31}^2 .~~~~~~
\nonumber
\eeqa
Using this and Eq.\eqref{eq:1105-4}, we derive the approximate formulas,
$\,\ep_m^{}\simeq -n_E^{}/(1\!-n_E^{})\,$ and
$\,\Delta_{m}^{} \simeq (1\!- n_E^{})\Delta\,$,\, by dropping small
$\,\mathcal{O}(s_x^2)\,$ terms. This just reproduces Eq.\eqref{eq:0111-1}
in the main text.

\vspace*{1mm}

In summary, we have proven the approximate formulas
\eqref{eq:0111}-\eqref{eq:0111-1} in the main text.

\vspace*{-1mm}
\noindent
\section{\hspace*{0mm}$\textrm{Further Tests of Matter Formula (23)}$}
\label{app:B}

In this Appendix, we further present two important tests of our new LUT
formula \eqref{eq:1023-4} by using the long baseline oscillation experiments
T2K\,\cite{T2K-PRD} and DUNE\,\cite{DUNE}.

\vspace*{1mm}

\begin{figure}[t]
\centering
\includegraphics[width=8.2cm,height=5.5cm]{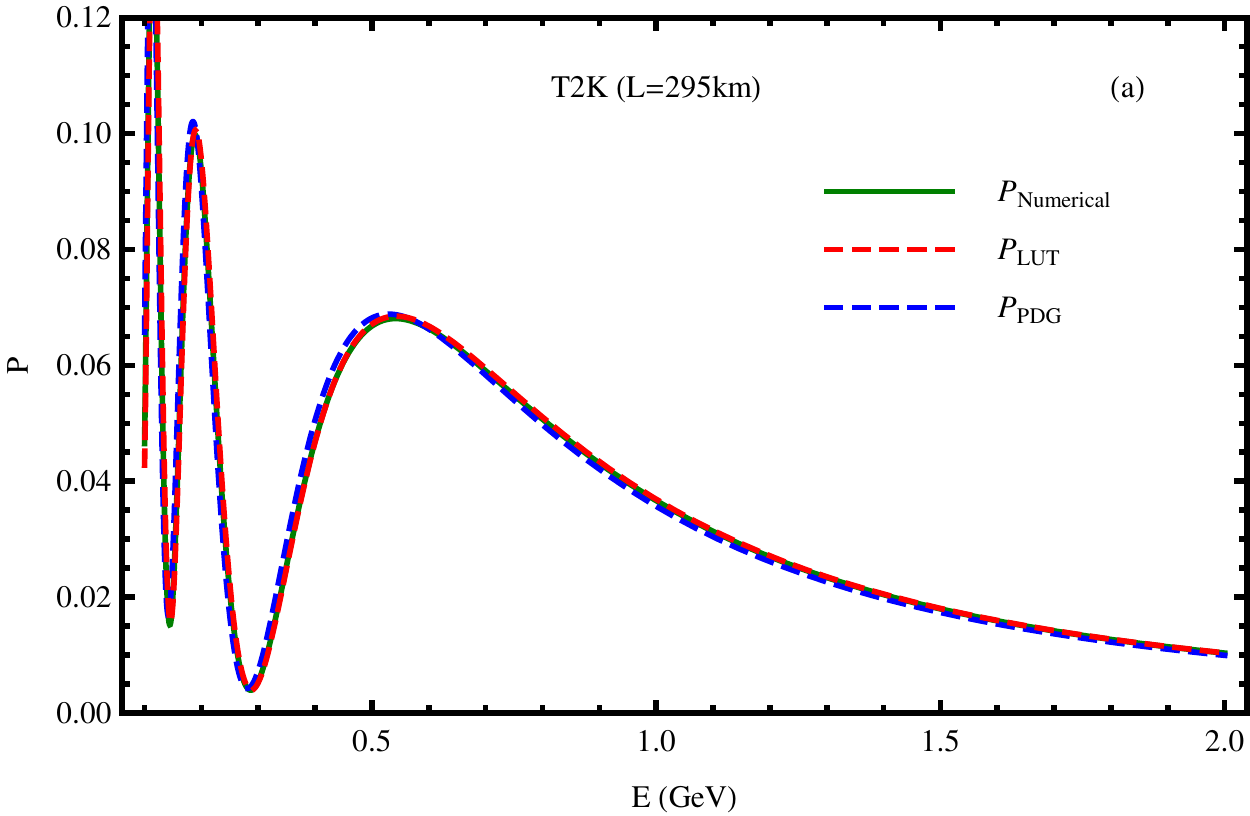}
\\
\hspace*{-3.4mm}
\includegraphics[width=8.5cm,height=5.5cm]{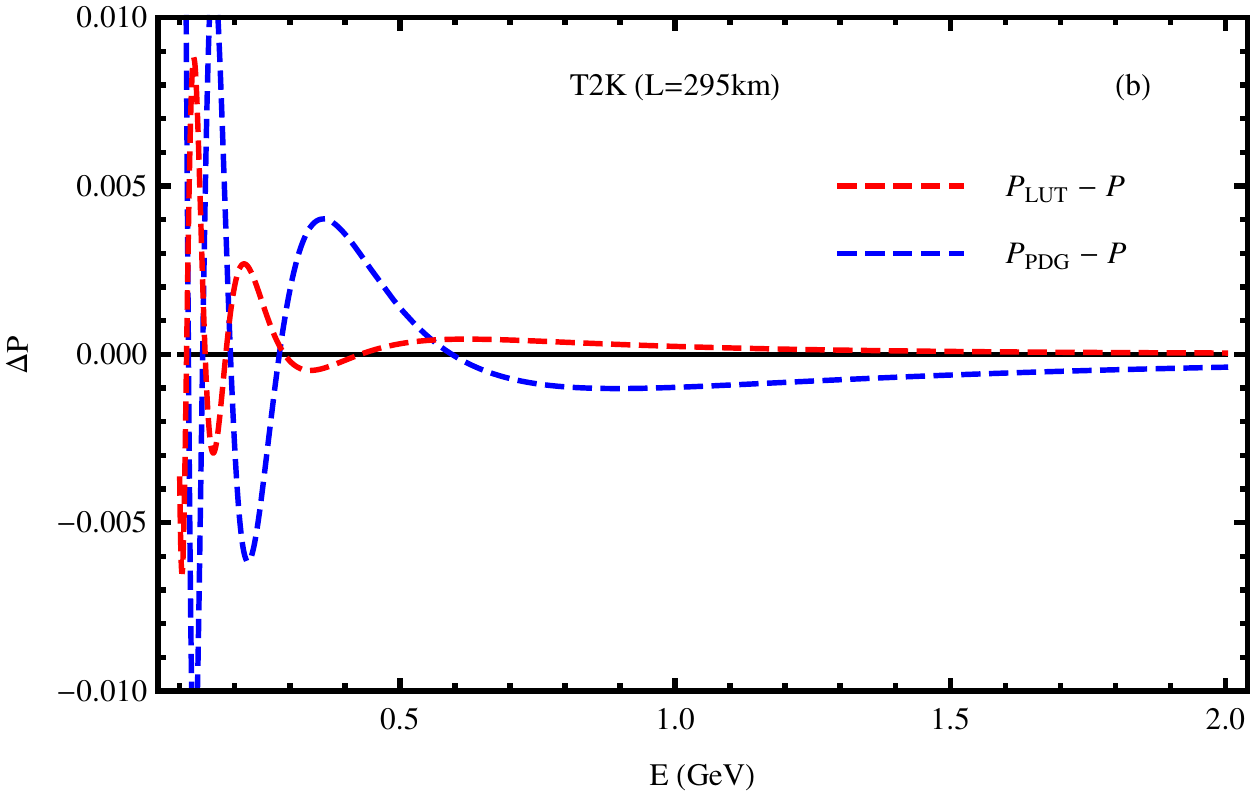}
\vspace*{-1mm}
\protect\caption{Same as Fig.\,\ref{fig:3} in the main text,
except changing the baseline length to $L= 295$\,km,
representing the case of T2K experiment.}
\label{fig:4}
\end{figure}

\begin{figure}[h]
\centering
\includegraphics[width=8.2cm,height=5.5cm]{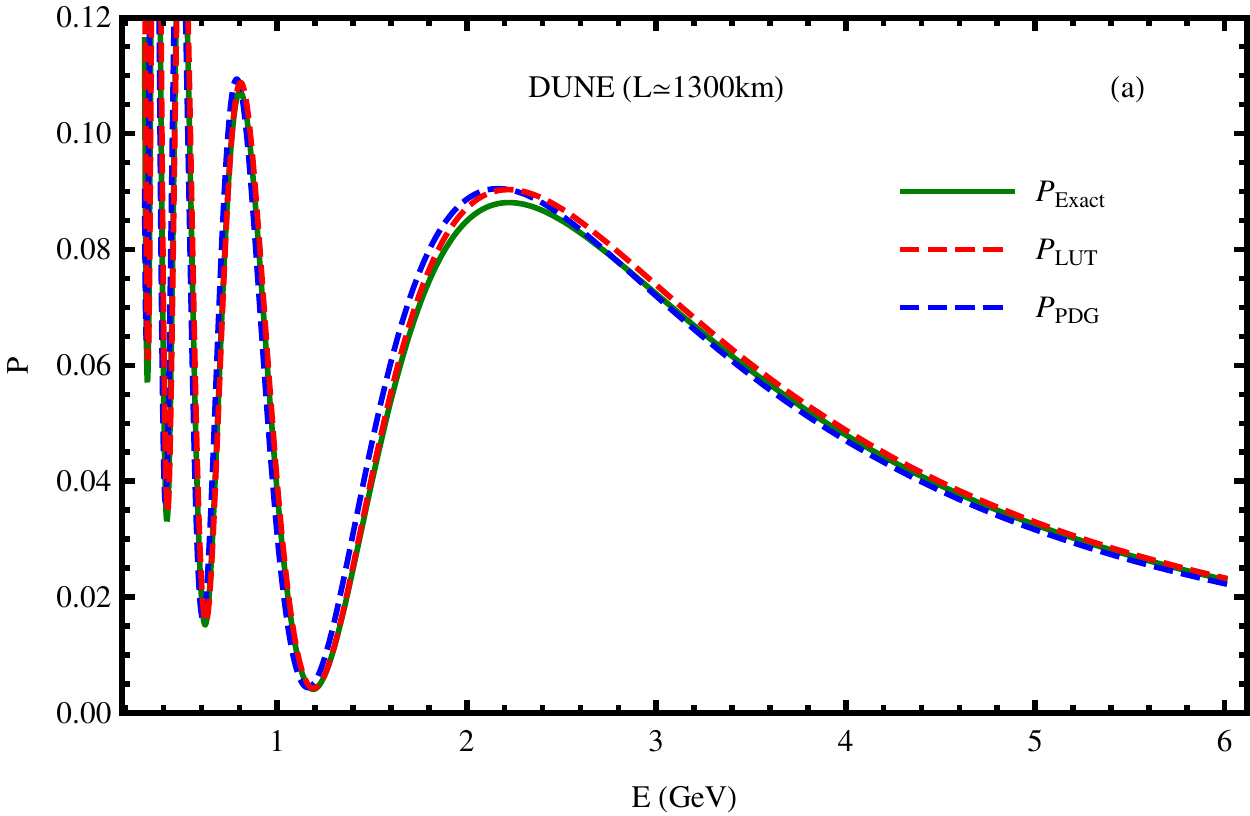}
\\
\hspace*{-3.4mm}
\includegraphics[width=8.5cm,height=5.5cm]{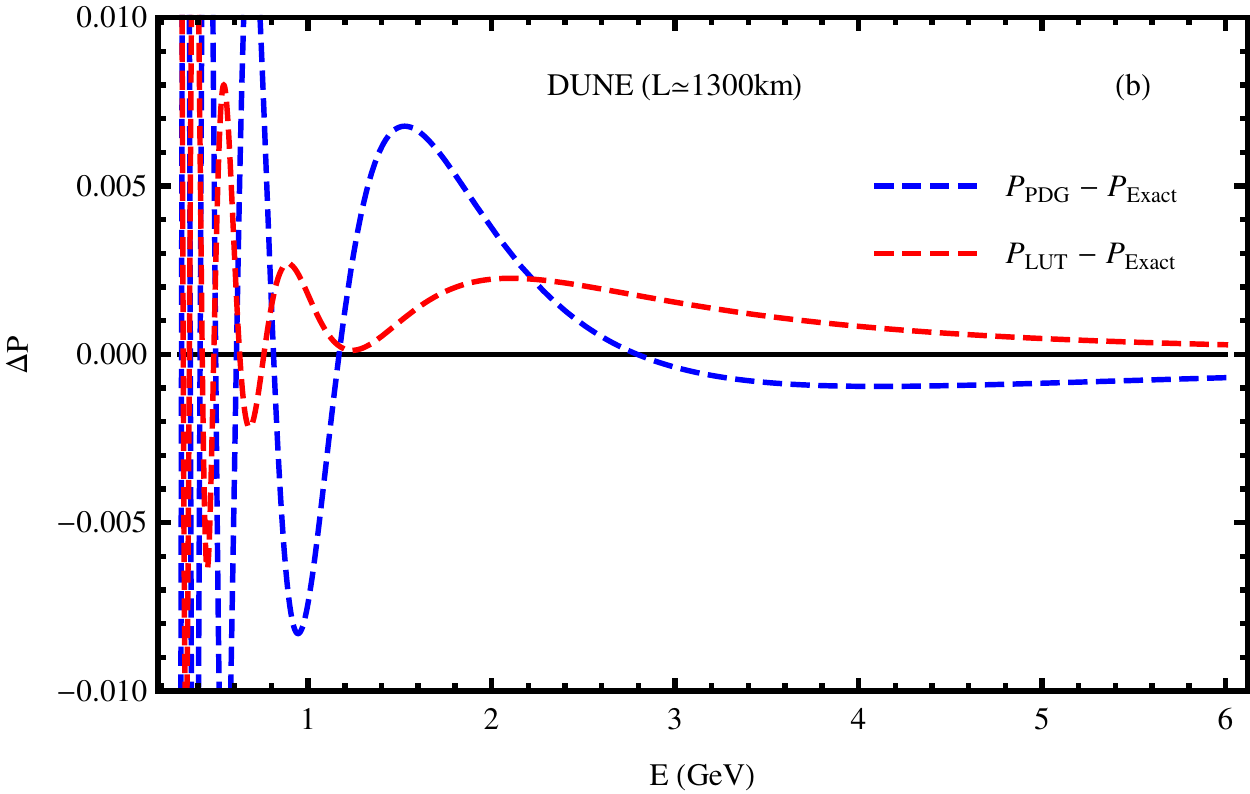}
\vspace*{-1mm}
\protect\caption{Same as Fig.\,3 in the main text, except changing the baseline length
to $L= 1300$\,km, representing the case of DUNE experiment.}
\label{fig:5}
\end{figure}

The baselines of the T2K and DUNE experiments
are $L=295$\,km and $L=1300$\,km, respectively.
We present the predictions of our Eq.\eqref{eq:LUTm-mue}
for T2K experiment in Fig.\,\ref{fig:4}(a) and for DUNE experiment
in Fig.\,\ref{fig:5}(a), by the red dashed curves.
Then, we compare them with the exact numerical results (green solid curves) in each plot.
For comparison, we further show the results of
the conventional formula \eqref{eq:PDGm-mue}
(used by the PDG \cite{PDG2014}) in the blue dashed curves.
We see that in each case, the three curves agree with each other to high precision,
similar to our findings in Fig.\,\ref{fig:3} for NO$\nu$A experiment.

\vspace*{1mm}

In Fig.\,\ref{fig:4}(b) and Fig.\,\ref{fig:5}(b),
we further compare the differences,
$\,\Delta P\!=\!P_{\text{LUT}}^{}-P\,$ (red dashed curves) and
$\,\Delta P=P_{\text{PDG}}^{}\!-\!P\,$ (blue dashed curves).
Again, these comparisons explicitly demonstrate that
our LUT formula \eqref{eq:LUTm-mue} is as accurate as (or better than) the
conventional PDG formula \eqref{eq:PDGm-mue}.

\vspace*{3mm}
\noindent
{\bf Acknowledgments:}
\\[1.5mm]
We are grateful to Sheldon Glashow, John Ellis,
Manfred Lindner, and Jose Valle for valuable discussions.
We thank Eligio Lisi for valuable discussions
during and after his visit to Tsinghua HEP Center.
We thank Yu-Chen Wang and Zhe Wang for useful discussions.
This research was supported in part by the National NSF of China
(under grants 11275101, 11135003, 11675086).

%

\end{document}